\documentclass[10pt]{iopart}
\usepackage{iopams,amssymb}

\def\unity{\mbox{\small 1} \!\! \mbox{1}}

\begin{document}
\title{Multi-Dimensional Hermite Polynomials in Quantum Optics}
\author{Pieter Kok and Samuel L.\ Braunstein}
\address{Informatics, University of Wales, Bangor, LL57 1UT, UK}

\begin{abstract}
 We study a class of optical circuits with vacuum input states consisting of
 Gaussian sources without coherent displacements such as down-converters and
 squeezers, together with detectors and passive interferometry (beam-splitters,
 polarisation rotations, phase-shifters etc.). We show that the {\em outgoing
 state} leaving the optical circuit can be expressed in terms of so-called
 multi-dimensional Hermite polynomials and give their recursion and
 orthogonality relations. We show how quantum teleportation of photon
 polarisation can be modelled using this description.
\end{abstract}

Suppose we have an optical circuit, that is, a collection of various optical 
components. It is usually important to know what the outgoing state of this 
circuit is. In this paper, we give a description of the outgoing state for a
special class of optical circuits with a special class of input states. 

First, in section \ref{circuit}, we define
this class of optical circuits and show that they can be described by 
so-called multi-dimensional Hermite polynomials. In section \ref{example}, we 
give an example of this description. Section \ref{hp} discusses the Hermite 
polynomials, and finally, in section \ref{sec:det}, we briefly consider the 
effect of imperfect detectors on the outgoing state.

\section{The Optical Circuit}\label{circuit}

What do we mean by an optical circuit? We can think of a black box with 
incoming and outgoing modes of the electro-magnetic field. The black
box transforms a state of the incoming modes into a (different) state of the
outgoing modes. The black box is what we call an {\em optical circuit}. We 
can now take a more detailed look inside the black box. We will consider three 
types of components.

First, the modes might be mixed by beam-splitters, or they may pick up a 
relative phase shift or polarisation rotation. These operations all belong  
to a class of optical components which preserve the photon number. We 
call them {\em passive} optical components. 

Secondly, we may find optical components such as lasers, down-converters or 
(optical) parametric amplifiers in the black box. These components can be 
viewed as photon sources, since they do not leave the photon number invariant. 
We will call these components {\em active} optical components. 

And finally, the box will generally include measurement devices, the outcomes
of which may modify optical components on the remaining modes depending on the 
detection outcomes. This is called {\em feed-forward} detection. We can 
immediately simplify optical circuits using feed-forward detection, by 
considering the family of fixed circuits
corresponding to the set of measurement outcomes (see also Ref.\ \cite{kok00}).
In addition, we can postpone the measurement to the end, where all the optical
components have `acted' on the modes. 

These three component types have their own characteristic mathematical 
description. A passive component yields a unitary evolution $U_i$, which can 
be written as 
\begin{equation}
 U_i = \exp\left( -i\kappa \sum_{jk} c_{jk} \hat{a}_j \hat{a}_k^{\dagger}
 - {\rm H.c.}\right)\; ,
\end{equation}
where H.c.\ denotes the Hermitian conjugate. This unitary evolution commutes 
with the total number operator $\hat{n}=\sum_j \hat{a}_j^{\dagger}\hat{a}_j$.

Active components also correspond to unitary transformations,
which can be written as $\exp(-itH_I^{(j)})$. Here $H_I^{(j)}$ is the 
interaction Hamiltonian associated with the $j^{\rm th}$ active component in a
sequence. 
This Hamiltonian does not necessarily commute with the total number operator. 
To make a typographical distinction between passive and active components, we 
denote the $i^{\rm th}$ passive component by $U_i$, and the $j^{\rm th}$ 
active component by its evolution in terms of the interaction Hamiltonian.

The mathematical description of the (ideal) measurement will correspond to 
taking the inner product of the outgoing state prior to the measurement with 
the eigenstate corresponding to the measurement. 

Now that we have the components of an optical circuit of $N$ modes, we have to 
combine them into an actual circuit. Mathematically, this corresponds to 
applying the unitary evolutions of
the successive components to the input state. Let $|\psi_{\rm in}\rangle$ be 
the input state and $|\psi_{\rm prior}\rangle$ the output state {\em prior} to 
the measurement. We then have (with $K>0$ some integer)
\begin{equation}\label{stateprior}
 |\psi_{\rm prior}\rangle = U_K e^{-it H_I^{(K)}} \ldots
 U_1 e^{-it H_I^{(1)}} U_0 |\psi_{\rm in}\rangle\; ,
\end{equation}
where it should be noted that $U_i$ might be the identity operator $\unity$
or a product of unitary transformations corresponding to passive components:
\begin{equation}
 U_i = \prod_k U_{i,k}\; .
\end{equation}
When the (multi-mode) eigenstate corresponding to the measurement outcome for 
a limited set of modes labelled $1,\ldots,M$ with $M<N$ is given by 
$|\gamma\rangle=|n_1,n_2,\ldots,n_M\rangle$ with $M$ the number of detected 
modes out of a total of $N$ modes, and $n_i$ the number of photons found in 
mode $i$, the state leaving the optical circuit in the undetected modes is 
given by 
\begin{equation}
 |\psi_{\rm out}\rangle_{M+1,\ldots,N} = _{1,\ldots,M}\langle\gamma|\psi_{\rm 
 prior}\rangle_{1,\ldots,N}\; .
\end{equation}

In this paper, we study the outgoing states $|\psi_{\rm out}\rangle$ for a 
special class of optical circuits. First, we assume that the input state is 
the vacuum on all modes. Thus, we effectively study optical circuits as state 
preparation devices. Secondly, our class of optical circuits include all 
possible passive components, but {\em only} active components with quadratic
interaction Hamiltonians:
\begin{equation}
 H_I^{(j)} = \sum_{kl} \hat{a}_k^{\dagger} R^{(j)}_{kl} \hat{a}_l^{\dagger} +
 \sum_{kl} \hat{a}_k R^{(j)*}_{kl} \hat{a}_l \; ,	
\end{equation}
where $R^{(j)}$ is some complex symmetric matrix. This matrix determines the 
behaviour of the $j^{\rm th}$ active component, which can be any combination 
of down-converters and squeezers. Finally, we consider ideal photo-detection, 
where the eigenstate corresponding to the measurement outcome can be written as
$|\gamma\rangle = |n_1,\ldots,n_M\rangle$.

The class of optical circuits we consider here is not the most general class, 
but it still includes important experiments like quantum teleportation 
\cite{bouwmeester97}, entanglement swapping \cite{pan98} and the demonstration
of GHZ correlations \cite{bouwmeester99b}. In section {\ref{example}} we show
how teleportation can be modelled using the methods presented here.

The state $|\psi\rangle$ prior to the photo-detection can be written 
in terms of the components of the optical circuit as
\begin{equation}\label{incoming}
 |\psi\rangle = U_K e^{-it {\mathcal{H}}_I^{(K)}} \ldots U_1 
 e^{-it H_I^{(1)}} |0\rangle\; .
\end{equation}
The creation and annihilation operators $\hat{a}_i^{\dagger}$ and $\hat{a}_i$ 
for mode $i$ satisfy the standard canonical commutation relations
\begin{equation}\label{comm}
 [\hat{a}_i,\hat{a}_j^{\dagger}] = \delta_{ij} \quad\mbox{and}\quad 
 [\hat{a}_i,\hat{a}_j] = [\hat{a}_i^{\dagger},\hat{a}_j^{\dagger}] = 0\; ,
\end{equation}
with $i,j=1\ldots N$. 

For any unitary evolution $U$, we have the relation
\begin{equation}
 U e^{R} U^{\dagger} = \sum_{l=0}^{\infty} \frac{U R^l U^{\dagger}}{l!} =
 \sum_{l=0}^{\infty} \frac{(U R U^{\dagger})^l}{l!} = e^{U R U^{\dagger}}\; .
\end{equation}
Furthermore, if $U$ is due to a collection of only passive components, such an 
evolution leaves the vacuum invariant: $U|0\rangle = |0\rangle$. Using these 
two properties it can be shown that Eq.\ (\ref{incoming}) can be written as 
\begin{equation}\label{in2}
 |\psi\rangle = \exp\left[ -\frac{1}{2} \sum_{i,j=1}^N \hat{a}^{\dagger}_i 
 A^{\dagger}_{ij} \hat{a}^{\dagger}_j + \frac{1}{2} \sum_{i,j=1}^N \hat{a}_i 
 A_{ij} \hat{a}_j \right] |0\rangle\; ,
\end{equation}
where $A$ is some complex symmetric matrix. We will now simplify this 
expression by normal-ordering this evolution.

Define $(\vec{a},A\vec{a}) \equiv \sum_{ij} \hat{a}_i A_{ij} \hat{a}_j$. As 
shown by Braunstein \cite{braunstein99}, we can rewrite Eq.\ (\ref{in2}) using 
two passive unitary transformations $U$ and $V$ as:
\begin{equation}\label{lambda}
 |\psi\rangle = U e^{-\frac{1}{2}(\vec{a}^{\dagger},\Lambda^{\dagger}
 \vec{a}^{\dagger}) + \frac{1}{2}(\vec{a},\Lambda\vec{a})} V^T |0\rangle\; ,
\end{equation}
where $\Lambda$ is a diagonal matrix with real non-negative eigenvalues 
$\lambda_i$.
This means that, starting from vacuum, the class of optical circuits we 
consider here is equivalent to a set of single-mode squeezers, followed by a 
passive unitary transformation $U$ and photo-detection. Since $\Lambda$ is 
diagonal, we can write Eq.~(\ref{lambda}) as 
\begin{equation}
 |\psi\rangle = U \left( \prod_{i=1}^N \exp\left[-\frac{\lambda^*_i}{2} 
 (\hat{a}_i^{\dagger})^2 + \frac{\lambda_i}{2} \hat{a}_i^2\right]\right) 
 |0\rangle\; .
\end{equation}
We can now determine the normal ordering of every factor $\exp[-
\frac{\lambda^*_i}{2} (\hat{a}_i^{\dagger})^2 + \frac{\lambda_i}{2} 
\hat{a}_i^2]$ separately. Note that the operators $\hat{a}_i^2$, 
$(\hat{a}_i^{\dagger})^2$ and $2\hat{a}_i^{\dagger}\hat{a}_i + 1$ generate an 
su(1,1) algebra. According to Refs.\ \cite{yuen76,fisher84,truax85}, this 
may be normal-ordered as
\begin{eqnarray}
 e^{-\frac{\lambda^*_i}{2} (\hat{a}_i^{\dagger})^2 + \frac{\lambda_i}{2} 
 \hat{a}_i^2} \cr \qquad\qquad =
 e^{-\hat{\lambda_i}^* \tanh|\frac{\lambda_i}{2}| (\hat{a}_i^{\dagger})^2}
 e^{-2 \ln (\cosh|\frac{\lambda_i}{2}|) \hat{a}_i^{\dagger} \hat{a}_i}
 e^{\hat{\lambda_i} \tanh|\frac{\lambda_i}{2}| \hat{a}_i^2}\; ,
\end{eqnarray}
where $\hat{\lambda_i} = \lambda_i / |\lambda_i|$. In general, when $L_{\pm}$
and $L_0$ are generators of an $su(1,1)$ algebra (i.e., when $A$ is unitary) we
find \cite{truax85}
\begin{equation}
 e^{-\frac{1}{2}(\tau L_+ \tau^* L_-)} = e^{-\hat{\tau}\tanh|\tau| L_+}
 e^{-2 \ln (\cosh|\tau|) L_0} e^{\hat{\tau}\tanh|\tau| L_-}\; ,
\end{equation}
with $\tau$ a complex coupling constant and $\hat{\tau}$ its orientation in 
the complex plane. When we now apply this operator to the vacuum, the 
annihilation operators
will vanish, leaving only the exponential function of the creation operators.
We thus have 
\begin{equation}\label{in3}
 |\psi\rangle = U e^{-\frac{1}{2}(\vec{a}^\dagger,\Lambda^*\vec{a}^{\dagger})} 
 V^T |0\rangle = e^{-\frac{1}{2}(\vec{a}^\dagger, B\vec{a}^{\dagger})}
 |0\rangle\; ,
\end{equation}
with $B \equiv U \Lambda^* U^{\dagger}$, again by virtue of the invariance
property of the vacuum. This is the state of the interferometer prior to
photo-detection. It corresponds to {\em multi-mode squeezed vacuum}.

The photo-detection itself can be modelled by successive application of 
annihilation operators. Every annihilation operator $\hat{a}_i$ removes a 
photon in 
mode $i$ from the state $|\psi\rangle$. Suppose the optical circuit employs 
$N$ distinct modes. We will now detect $M$ modes, finding $n_1+\ldots +n_M = 
N_{\rm tot}$ photons (with $M<N$). These modes can be relabelled 1 to $M$. 
The vector $\vec{n}$ denotes the particular detector `signature': $\vec{n} = 
(n_1,\ldots,n_M)$ means that $n_1$ photons are detected in mode 1, $n_2$ in 
mode 2, and so on. The freely propagating outgoing state 
$|\psi_{\vec{n}}\rangle$ can then be described as
\begin{equation}\label{det}
 |\psi_{\vec{n}}\rangle = ~_{1..M}\langle n_1,\ldots, n_M | \psi\rangle_{1..N} 
 = c_{\vec{n}} \langle 0| \hat{a}_1^{n_1}\cdots \hat{a}_M^{n_M}|\psi\rangle\; .
\end{equation}
Here, $c_{\vec{n}} = (n_1!\cdots n_M!)^{-\frac{1}{2}}$.

At this point we find it convenient to introduce the $N$-mode Bargmann
representation \cite{bargmann}. The creation and annihilation operators obey 
the commutation relations given in Eq.\ (\ref{comm}). We can replace these 
operators with c-numbers and their derivatives according to
\begin{equation}
 \hat{a}_i^{\dagger} \rightarrow \alpha_i \quad\mbox{and}\quad 
 \hat{a}_i \rightarrow \partial_i \equiv \frac{\partial}{\partial\alpha_i}\; .
\end{equation}
The commutation relations then read
\begin{equation}
 [\partial_i,\alpha_j] = \delta_{ij} \quad\mbox{and}\quad
 [\partial_i,\partial_j] = [\alpha_i,\alpha_j] = 0\; .
\end{equation}
Note that the actual values of $\alpha_i$ are irrelevant (the creation and 
annihilation operators do not have numerical values either); what matters here 
is the functional relationship between $\alpha_i$ and $\partial_{\alpha_i}$.

The state created by the optical circuit in this representation (prior to the 
detections, analogous to Eq.\ (\ref{in3})) in the Bargmann representation is
\begin{equation}
 \psi(\vec\alpha) = \exp\left[ -\frac{1}{2}(\vec\alpha,B\vec\alpha)\right]
 = \exp\left[ -\frac{1}{2}\sum_{ij}\alpha_i B_{ij}\alpha_j \right]\; .
\end{equation}
Returning to Eq.\ (\ref{det}), we can write the freely propagating state 
after detection of the auxiliary modes in the Bargmann representation as
\begin{equation}
 \psi_{\vec{n}}(\vec\alpha) \propto c_{\vec{n}}\; \partial_1^{n_1} \cdots 
 \partial_M^{n_M} \left. e^{-\frac{1}{2}(\vec\alpha,B\vec\alpha)} 
 \right|_{\vec\alpha'=0}\; ,
\end{equation}
up to some normalisation factor, where $\vec\alpha' = (\alpha_1,\ldots,
\alpha_M)$. By setting $\vec\alpha'=0$ we ensure that {\em no more} that 
$n_i$ photons are present in mode $i$. It plays the role of the vacuum bra 
in Eq.\ (\ref{det}).

Now that we have an expression for the freely propagating state emerging
from our optical setup after detection, we seek to simplify it. We can 
multiply $\psi_{\vec{n}}(\vec\alpha)$ by the identity operator $I$, written as
\begin{equation}
 I = (-1)^{2N_{\rm tot}} \exp\left[-\frac{1}{2}(\vec\alpha,B\vec\alpha)\right]
 \; \exp\left[\frac{1}{2}(\vec\alpha,B\vec\alpha)\right]\; ,
\end{equation}
where $N_{\rm tot}$ is the total number of detected photons. We then find the
following expression for the unnormalised freely propagating state created by 
our optical circuit:
\begin{equation}\label{out}
 \psi_{\vec{n}}(\vec\alpha) \propto \left. c_{\vec{n}} (-1)^{N_{\rm tot}} 
 H_{\vec{n}}^B (\vec\alpha)\; e^{-\frac{1}{2}(\vec\alpha,B\vec\alpha)}
 \right|_{\vec\alpha'=0}\; .
\end{equation}

Now we introduce the so-called {\em multi-dimensional} Hermite polynomial,
or MDHP for short:
\begin{equation}\label{mdhp}
 H_{\vec{n}}^B (\vec\alpha) = (-1)^{N_{\rm tot}} e^{\frac{1}{2}(\vec\alpha,
 B\vec\alpha)}\; \frac{\partial^{n_1}}{\partial\alpha_1^{n_1}} \cdots 
 \frac{\partial^{n_M}}{\partial\alpha_M^{n_M}}\; e^{-\frac{1}{2}
 (\vec\alpha,B\vec\alpha)}\; .
\end{equation}
The use of multi-dimensional Hermite polynomials and Hermite polynomials of 
two variables have previously been used to describe $N$-dimensional
first-order systems \cite{dodonov84,klauderer93} and photon statistics
\cite{vourdas87,dodonov94,kok99}. Here, we have shown that the lowest order of 
the {\em outgoing} state of optical circuits with quadratic components (as 
described by Eq.\ (\ref{in2})) and conditional photo-detection can be expressed
directly in terms of an MDHP.

In physical systems, the coupling constants (the $\lambda_i$'s) are usually
very small (i.e., $\lambda_i\ll 1$ or possibly $\lambda_i \lesssim 1$). This 
means that for all practical purposes only the first order term in 
Eq.~(\ref{out}) is important (i.e., for small $\lambda_i$'s we can approximate 
the exponential by 1). 
Consequently, studying the multi-dimensional Hermite polynomials yield
knowledge about the typical states we can produce using Gaussian sources 
without coherent displacements. In section \ref{hp} we take a closer look at 
these polynomials, but first we consider the description of quantum 
teleportation in this representation.

\section{Example: Quantum Teleportation}\label{example}

As an example of how to determine the outgoing state of an optical circuit,
consider the teleportation experiment by Bouwmeester {\em et al}.\ 
\cite{bouwmeester97}. The optical circuit corresponding to this experiment
consists of eight incoming modes, all in the vacuum state. Physically, there 
are four spatial modes $a$, $b$, $c$ and $d$, all with two polarisation 
components $x$ and $y$. Two down-converters create entangled polarisation 
states; they belong to the class of active Gaussian components without 
coherent displacements. Mode $a$ undergoes a polarisation rotation over an 
angle $\theta$ and modes $b$ and $c$ are mixed in a 50:50 beam-splitter.
Finally, modes $b$ and $c$ emerging from the beam-splitter are detected with 
polarisation insensitive detectors and mode $a$ is detected using a 
polarisation sensitive detector. The state which is to be teleported is 
therefore given by
\begin{equation}
 |\Psi\rangle = \cos\theta |x\rangle - \sin\theta |y\rangle\; .
\end{equation}

The state prior to the detection and normal ordering (corresponding to 
Eq.~(\ref{stateprior})) is given by ($\tau$ is a coupling constant) 
\begin{equation}
 |\psi_{\rm prior}\rangle = U_{\rm BS} U_{\theta} 
 e^{\tau (\vec{u}^{\dagger},L\vec{u}^{\dagger})/2 + \tau^* (\vec{u},L\vec{u})/2
 + \tau (\vec{v}^{\dagger},L\vec{v}^{\dagger})/2 + \tau^* (\vec{v},L\vec{v})/2}
 |0\rangle\; ,
\end{equation}
with 
\begin{equation}
 L = \frac{1}{\sqrt{2}} \left( 
 \begin{array}{rrrr}
  0 & 0 & 0 & 1 \\
  0 & 0 & 1 & 0 \\
  0 & 1 & 0 & 0 \\
  1 & 0 & 0 & 0 
 \end{array}
 \right)
\end{equation}
and $\vec{u}^{\dagger}=(\hat{a}^{\dagger}_x,\hat{a}^{\dagger}_y,
\hat{b}^{\dagger}_x,\hat{b}^{\dagger}_y)$, $\vec{v}^{\dagger}=
(\hat{c}^{\dagger}_x,\hat{c}^{\dagger}_y,\hat{d}^{\dagger}_x,
\hat{d}^{\dagger}_y)$. This can be written as
\begin{equation}\label{priorstate}
 |\psi_{\rm prior}\rangle = \exp\left[ \frac{\tau}{2} 
 (\vec{a}^{\dagger},A\vec{a}^{\dagger}) + \frac{\tau^*}{2} (\vec{a},A\vec{a}) 
 \right] |0\rangle\; ,
\end{equation}
with $\vec{a}\equiv(\hat{a}_x,\ldots,\hat{d}_y)$ and $A$ the (symmetric) matrix
\begin{equation}
 A = \frac{1}{\sqrt{2}} \left(
 \begin{array}{rrrrrrrr}
  0 & 0 & -\sin\theta & \cos\theta & -\sin\theta & \cos\theta & 0 & 0 \\ 
  & 0 & \cos\theta & \sin\theta & \cos\theta & \sin\theta & 0 & 0 \\ 
  & & 0 & 0 & 0 & 0 & 0 & -1 \\ 
  & & & 0 & 0 & 0 & -1 & 0 \\ 
  & & & & 0 & 0 & 0 & 1 \\ 
  & & & & & 0 & 1 & 0 \\ 
  & & & & & & 0 & 0 \\ 
  & & & & & & & 0 
 \end{array}
 \right)\; .
\end{equation}

We now have to find the normal ordering of Eq.~(\ref{priorstate}). Since $A$
is unitary, the polynomial $(\vec{a}^{\dagger},A\vec{a}^{\dagger})$ is a 
generator of an $su(1,1)$ algebra. According to Truax \cite{truax85}, the 
normal ordering of the exponential thus yields a state
\begin{equation}\label{telprior}
 |\psi_{\rm prior}\rangle = \exp\left[ \frac{\xi}{2}
 (\vec{a}^{\dagger},A\vec{a}^{\dagger}) \right] |0\rangle\; ,
\end{equation}
with $\xi = (\tau\tanh|\tau|)/|\tau|$. The lowest order contribution after 
three detected photons is due to the term $\xi^2(\vec{a}^{\dagger},
A\vec{a}^{\dagger})^2/8$. However, first we write Eq.~(\ref{telprior}) in the 
Bargmann representation:
\begin{equation}
 \psi_{\rm prior}(\vec{\alpha}) = \exp\left[ \frac{\xi}{2}
 (\vec{\alpha},A\vec{\alpha}) \right]\; ,
\end{equation}
where $\vec{\alpha}=(\alpha_{a_x},\ldots,\alpha_{d_y})$ and $\vec{\alpha}'=
(\alpha_{a_x},\ldots,\alpha_{c_y})$. The polarisation independent 
photo-detection is then modelled by the differentiation $(\partial_{b_x}
\partial_{c_y} - \partial_{b_y}\partial_{c_x})$. Given a detector hit in mode
$a_x$, the polarisation sensitive detection of mode $a$ is modelled by 
$\partial_{a_x}$:
\begin{equation}
 \psi_{\rm out}(\vec{\alpha}) = \left. \partial_{a_x}\left( \partial_{b_x}
 \partial_{c_y} - \partial_{b_y}\partial_{c_x} \right) \exp\left[ \frac{\xi}{2}
 (\vec{\alpha},A\vec{\alpha}) \right] \right|_{\vec{\alpha}'=0}\; .
\end{equation}
The outgoing state in the Bargmann representation is thus given by
\begin{equation}
 \psi_{\rm out}(\vec{\alpha}) = \left(\cos\theta\, \alpha_{d_x} + \sin\theta\,
 \alpha_{d_y}\right) e^{\frac{\xi}{2} (\vec{\alpha},A\vec{\alpha})}\; ,
\end{equation}
which is the state teleported from mode $a$ to mode $d$ in the Bargmann 
representation. This procedure 
essentially amounts to evaluating the multi-dimensional Hermite polynomial
$H_{\vec{n}}^A(\vec{\alpha})$. Note that the polarisation independent 
detection of modes $b$ and $c$ yield a {\em superposition} of the MDHP's.

\section{The Hermite Polynomials}\label{hp}

The one-dimensional Hermite polynomials are of course well known from the
description of the linear harmonic oscillator in quantum mechanics. These
polynomials may be obtained from a generating function $G$. Furthermore, 
there exist two recursion relations and an orthogonality relation between 
them. The theory of multi-dimensional Hermite polynomials with real variables 
has been developed by Appell and Kemp\'e de F\'eriet \cite{appell26} and in 
the Bateman project \cite{bateman}. Mizrahi derived an expression for real 
MDHP's from an $n$-dimensional generalisation of the Rodriguez formula 
\cite{mizrahi75}. We will now give the generating function for the complex 
MDHP's given by Eq.\ (\ref{mdhp}) and consecutively derive the recursion 
relations and the orthogonality relation (see also Ref.\ \cite{klauderer93}). 

Define the generating function $G_B(\vec{\alpha},\vec{\beta})$ to be
\begin{equation}\label{genfunc}
 G_B(\vec{\alpha},\vec{\beta}) = e^{(\vec{\alpha},B\vec{\beta}) -
 \frac{1}{2} (\vec{\beta},B\vec{\beta})} = \sum_{\vec{n}} 
 \frac{\beta_1^{n_1}}{n_1!} \cdots \frac{\beta_M^{n_M}}{n_M!}
 H^B_{\vec{n}} (\vec{\alpha})\; .
\end{equation}
$G_B(\vec{\alpha},\vec{\beta})$ gives rise to the MDHP in 
Eq.\ (\ref{mdhp}), which determines this particular choice. Note that the
inner product $(\vec{\alpha},B\vec{\beta})$ does not involve any complex 
conjugation. If complex conjugation was involved, we would have obtained 
different polynomials (which we could also have called multi-dimensional 
Hermite polynomials, but they would not bear the same relationship to optical
circuits). 

In the rest of the paper we use the following notation: by $\vec{n}-e_j$ we
mean that the $j^{\rm th}$ entry of the vector $\vec{n}=(n_1,\ldots,n_M)$ is
lowered by one, thus becoming $n_j-1$. By differentiation of both sides of 
the generating function in Eq.\ (\ref{genfunc}) we can thus show that the 
first recursion relation becomes
\begin{equation}
 \frac{\partial}{\partial\alpha_i} H^B_{\vec{n}} (\vec{\alpha}) = 
 \sum_{j=1}^M B_{ij} n_j H^B_{\vec{n}-e_j} (\vec{\alpha})\; .
\end{equation}

The second recursion relation is given by
\begin{equation}
 H_{\vec{n}+e_i}^B (\vec{\alpha}) - \sum_{j=1}^M B_{ij}\alpha_j H_{\vec{n}}^B
 (\vec{\alpha})+\sum_{j=1}^M B_{ij} n_j H_{\vec{n}-e_j}^B (\vec{\alpha}) =0\; ,
\end{equation}
which can be proved by mathematical induction using
\begin{equation}
 \sum_{k=1}^M B_{ik} n_k H_{\vec{n}-e_k+e_i}^B (\vec{\alpha}) - B_{ii}
 H_{\vec{n}}^B (\vec{\alpha}) = \sum_{k=1}^M B_{ik} m_k H_{\vec{m}+e_i}^B
 (\vec{\alpha})\; .
\end{equation}
Here, we have chosen $\vec{m}=\vec{n}-e_k$.

The orthogonality relation is somewhat more involved. Ultimately, we want 
to use this relation to determine the normalisation constant of the states
given by Eq.\ (\ref{out}). To find this normalisation we have to evaluate the 
integral
\begin{equation}\nonumber
 \int_{\mathbb{C}^N} d\vec\alpha\, \psi_{\vec{n}}^* (\vec\alpha) \psi_{\vec{m}}
 (\vec\alpha)\; .
\end{equation}
The state $\psi_{\vec{n}}$ includes $|_{\vec\alpha'=0}$, which translates into
a delta-function $\delta(\vec\alpha')$ in the integrand. The relevant 
integral thus becomes
\begin{equation}\nonumber
 \int_{\mathbb{C}^N} d\vec\alpha\, e^{-{\rm Re}(\vec\alpha,B\vec\alpha)}
 \left[ H_{\vec{n}}^B (\vec\alpha) \right]^* H_{\vec{m}}^B (\vec\alpha)\,
 \delta(\vec\alpha')\; .
\end{equation}
From the orthonormality of different quantum states we know that this 
integral must be proportional to $\delta_{\vec{n},\vec{m}}$. 

Since in the Bargmann representation we are only concerned with the {\em 
functional} relationship between $\alpha_i$ and $\partial_{\alpha_i}$ and not
the actual values, we can choose $\alpha_i$ to be real. To stress this, we
write $\alpha_i\rightarrow x_i$. The orthogonality relation is thus derived 
from 
\begin{equation}
 \int_{\mathbb{R}^N} d\vec{x}\, \psi^*_{\vec{n}}(\vec{x})
 \psi_{\vec{m}}(\vec{x}) = \int_{\mathbb{R}^N} d\vec{x}\,
 e^{-(\vec{x},{\rm Re}(B)\vec{x})} H_{\vec{n}}^{B^*} 
 (\vec{x}) H_{\vec{m}}^B (\vec{x})\, \delta(\vec{x}') \; .
\end{equation}
where $\delta(\vec{x}')$ is the real version of $\delta(\vec\alpha')$. 
Following Klauderer \cite{klauderer93} we find that
\begin{eqnarray}
 \int d\vec{x}\, e^{-(\vec{x},{\rm Re}(B)\vec{x})}
 H_{\vec{n}}^{B^*} (\vec{x}) H_{\vec{m}}^B (\vec{x}) = \cr \qquad
 (-1)^{N_{\rm tot}}\, \int d\vec{x}\, e^{-\frac{1}{2}(\vec{x},B\vec{x})}
 \partial_{\vec{x}}^{\vec{n}}\, 
 \left[ e^{-\frac{1}{2}(\vec{x},B^*\vec{x})}\right] H_{\vec{m}}^B
 (\vec{x})\; ,
\end{eqnarray}
where $\partial_{\vec{x}}^{\vec{n}}$ is the differential operator 
$\partial_{x_1}^{n_1}\cdots\partial_{x_M}^{n_M}$ acting solely on the 
exponential function. We now integrate the right-hand side by parts, yielding
\begin{eqnarray}
 (-1)^{N_{\rm tot}}\, \int d\vec{x}\, e^{-\frac{1}{2}(\vec{x},B\vec{x})}
 \partial_{\vec{x}}^{\vec{n}}\, 
 e^{-\frac{1}{2}(\vec{x},B^*\vec{x})} H_{\vec{m}}^B (\vec{x})=\cr\qquad
 \left.(-1)^{N_{\rm tot}}\, \int d'\vec{x}\, e^{-\frac{1}{2}(\vec{x},B\vec{x})}
 \partial_{\vec{x}}^{\vec{n}-e_i}\,
 e^{-\frac{1}{2}(\vec{x},B^*\vec{x})} H_{\vec{m}}^B (\vec{x})
 \right|^{+\infty}_{x_i = -\infty} \cr\qquad - (-1)^{N_{\rm tot}}\, \int 
 d\vec{x}\, e^{-\frac{1}{2}(\vec{x},B\vec{x})}
 \partial_{\vec{x}}^{\vec{n}-e_i}\, e^{-\frac{1}{2}
 (\vec{x},B^*\vec{x})} \partial_{x_i} H_{\vec{m}}^B (\vec{x})\; ,
\end{eqnarray}
with $d'\vec{x}=dx_1\cdots dx_{i-1} dx_{i+1}\cdots dx_{N}$.
The left-hand term is equal to zero when ${\rm Re}(B)$ is positive definite,
i.e., when $(\vec{x},{\rm Re}(B)\vec{x})>0$ for all non-zero $\vec{x}$. 
Repeating this procedure $n_i$ times yields
\begin{eqnarray}
 \int d\vec{x}\, e^{-(\vec{x},{\rm Re}(B)\vec{x})}
 H_{\vec{n}}^{B^*} (\vec{x}) H_{\vec{m}}^B (\vec{x}) = \cr \qquad
 (-1)^{N_{\rm tot}+n_i}\, \int d\vec{x}\, e^{-\frac{1}{2}(\vec{x},B\vec{x})}
 \partial_{\vec{x}}^{\vec{n}-n_ie_i}\,
 e^{-\frac{1}{2}(\vec{x},B^*\vec{x})} \partial_{x_i}^{n_i} 
 H_{\vec{m}}^B (\vec{x})\; .
\end{eqnarray}
When there is at least one $n_i>m_i$, differentiating the MDHP $n_i$ times to
$x_i$ will yield zero. Thus we have 
\begin{equation}
 \int d\vec{x}\, e^{-(\vec{x},{\rm Re}(B)\vec{x})} 
 H_{\vec{n}}^{B^*} (\vec{x}) H_{\vec{m}}^B (\vec{x}) = 0 \quad\mbox{for}~
 \vec{n}\neq\vec{m}
\end{equation}
when ${\rm Re}(B)$ is positive definite and $n_i \neq m_i$ for any $i$. The
case where $\vec{n}$ equals $\vec{m}$ is given by
\begin{equation}
 \int d\vec{x}\, e^{-\frac{1}{2}(\vec{x},{\rm Re}(B)\vec{x})} 
 H_{\vec{n}}^{B^*} (\vec{x}) H_{\vec{m}}^B (\vec{x}) = \delta_{\vec{n}\vec{m}}
 {\mathcal{N}}\; ,
\end{equation}
where $\delta_{\vec{n}\vec{m}}$ denotes the product of $\delta_{n_i m_i}$ with
$1\leq i \leq N$. Here, $\mathcal{N}$ is equal to
\begin{equation}
 {\mathcal{N}} \equiv 2^{N_{\rm tot}}\, B_{11}^{n_1} \cdots B_{NN}^{n_N}\, 
 n_1!\cdots n_N! \, \left|\pi^{-1} B\right|^{-\frac{1}{2}}\; .
\end{equation}
For the proof of this identity we refer to Ref.\ \cite{klauderer93}. 

\section{Imperfect Detectors}\label{sec:det}

So far, we only considered the use of ideal photo-detection. That is, we 
assumed that the detectors tell us exactly and with unit efficiency how many 
photons were present in the detected mode. However, in reality such detectors
do not exist. In particular we have to incorporate losses (non-perfect
efficiency) and dark counts. Furthermore, we have to take into account the
fact that most detectors do not have a single-photon resolution (i.e., they
cannot distinguish a single photon from two photons) \cite{kok99}. 

This model is not suitable when we want to include dark counts. These
unwanted light sources provide thermal light, which is not of the form of 
Eq.\ (\ref{in2}). In single-shot experiments, however, dark counts can be
neglected when the detectors operate only within a narrow time interval.

We can model the efficiency of a detector by placing a beam splitter with 
transmission amplitude $\eta$ in front of a perfect detector \cite{kok99}. The 
part of the
signal which is reflected by the beam-splitter (and which will therefore never 
reach the detector) is the loss due to the imperfect detector. Since beam
splitters are part of the set of optical devices we allow, we can make this
generalisation without any problem. We now trace out all the reflected modes
(they are truly `lost'), and end up with a mixture in the remaining undetected 
modes.

Next, we can model the lack of single-photon resolution by using the 
relative probabilities $p(n|k)$ and $p(m|k)$ of the actual number $n$ or $m$ 
of detected photons conditioned on the indication of $k$ photons in the 
detector (as described in Ref.\ \cite{kok99}). We can determine the pure states
according to $n$ and $m$ detected photons, and add them with relative weights
$p(n|k)$ and $p(m|k)$. This method is trivially generalised for more than two 
possible detected photon numbers.

Finally, we should note that our description of this class of optical circuits
(in terms of multi-dimensional Hermite polynomials) is essentially a one-way
function. Given a certain setup, it is relatively straightforward to determine
the outgoing state of the circuit. The other way around, however, is very 
difficult. As exemplified by our efforts in Ref.\ \cite{kok00}, it is almost
impossible to obtain the matrix $B$ associated with an optical circuit which 
produces a particular predetermined state from a Gaussian source. 

\section{Conclusions}

In this paper, we have derived the general form of squeezed multi-mode vacuum 
states conditioned on photo-detection of some of the modes. To lowest order, 
the outgoing states in the Bargmann representation
are proportional to multi-dimensional Hermite polynomials. As an example, we
showed how teleportation can be described this way. 

\section*{Acknowledgements}

This work was supported in part under project QUICOV under the 
IST-FET-QIPC programme. PK would like to thank A.\ Vourdas for stimulating 
discussions.

\end{document}